\documentclass[12pt]{article}
\usepackage{epsfig}
\def\i2           {\mbox{$\frac{i}{2}$}}

\begin{document}
\title{ Second harmonic  generation in the moving media }
\author{ Mahboubeh Ghalandari \footnote {Electronic mail:
mahboubeh.ghalandari@gmail.com, ghalandari@qut.ac.ir, Tel: +98-9112076607}\\
{\small Department of Physics, Faculty of Sciences, Qom University
of Technology, Qom, Iran}\\
}
\date{}
\maketitle
\begin{abstract}
Because of the importance of second harmonic generation in some
nonlinear media, in this paper, we investigated induced second
harmonic generation in diamond where there is no intrinsic second
order susceptibility, $\chi^{(2)}$. The electric field is proposed
to introduce moving susceptibility of the second order and induce
second harmonic generation. Then, spatiotemporal (QPM) is applied to
optimize the induced second harmonic generation. Numerical results
reveals that in this way, the induced second harmonic is found at
the frequency of $\omega_{2}=2\omega_{0}\mp\bigtriangleup\omega$
rather than $\omega_{0}$.
\end{abstract}
\newpage
\vskip 1cm {\bf\large 1. Introduction} \vskip 0.5cm \ Recent
progressions in the understanding of light propagation have led to
proof that by using intense laser pulses propagating in a nonlinear
medium, it is possible to create an effective medium that flows with
the same speed as the laser pulse, i.e. at speeds close to or, as a
consequence of dispersion, even higher than the speed of light at
other frequencies in the medium [1-5]. In other words, intense pulse
of light will excite a polarization wave in nonlinear medium, i.e. a
pulsed change in the refractive index that travels locked to the
light pulse. This refractive index change is therefore traveling at
the speed of light in the medium. It is possible to therefore to
create not only flowing media but also, for example, rotating media,
oscillating or accelerating media. These models may be used to study
how quantum field theories in these curved space times and study
various effects such as the dynamical Casimir effect and
acceleration radiation [1-5]. Investigating light in moving media we
recently (re)discovered the important role of negative frequencies
and more in general, of the negative frequency content of optical
waves in nonlinear optics [6,7]. In this paper, we use the moving
media for second harmonic generation in media such as diamond where
there is no second order susceptibility,  (2). One possible
experiment related to this work can be a CO2 CW laser (10 micron
wavelength) propagating with a standard near-infrared probe pulse
(e.g. 800 nm wavelength), so that the 10 micron wavelength laser
creates the varying effective  and thus converts the 800 nm to its
second harmonic. A possible material to do this could be any
material that is transparent from the UV to above 10 microns, e.g.
diamond.
\\
This paper is organized as follows: Theoretical foundations is
introduced in section 2. In section 3, simulations are presented.
Results and discussions is introduced in section 4. Finally, in
section 5, conclusions, will be given.
 \vskip 1cm {\bf\large 2.
Theoretical foundations} \vskip 0.5cm \ Let us consider the light
polarized in the y direction and propagating in the z direction. The
curl Maxwell equations are
\begin{equation}
\mathrm{\frac{\partial E_{y}}{\partial z}=\mu_{0}\frac{\partial
H_{x}}{\partial t}},
\end{equation}
\begin{equation}
\mathrm{\frac{\partial H_{x}}{\partial z}=\epsilon_{0}\frac{\partial
D_{y }}{\partial t}},
\end{equation}
We do pseudospectral space-domain(PSSD) simulations [8] that solve
Maxwell equations without extra approximation. We use PSSD procedure
instead of finite-difference time-domain (FDTD) [9] and
pseudospectral time-domain (PSTD) [10]. Because PSSD approach
provides greater freedom for accurate modelling of dispersion and
nonlinear processes. This is due to description of time integrals by
convolutions in time at a particular point in space. In the case of
linear dispersion and non-instantaneous nonlinear response, the
electric field displacement D, at a particular point in space, is
given by
\begin{equation}
\mathrm{D(t)=\int_{-\infty}^\infty d\acute{t} \tilde{\epsilon}^{(1)}
(t-\acute{t})
E(\acute{t})+\int_{-\infty}^\infty\int_{-\infty}^\infty d\acute{t}
d\acute{\acute{t}} \tilde{\epsilon}^{(2)}(t-\acute{t},
t-\acute{\acute{t}}) E(\acute{t}) E(\acute{\acute{t}})+...},
\end{equation}
Where $\tilde{\epsilon}^{(1)}$ is the first-order linear
permittivity, and $\tilde{\epsilon}^{(2)}$ is the second-order
nonlinear permittivity. By using of Born-Oppenheimer approximation
[11], the double integral in eq.(3) reduces to a single convolution
in time
\begin{equation}
\mathrm{D(t)=\int_{-\infty}^\infty d\acute{t} \tilde{\epsilon}^{(1)}
(t-\acute{t}) E(\acute{t})+ E(t) \int_{-\infty}^\infty d\acute{t}
\tilde{\epsilon}^{(2)}_{BO}(t-\acute{t}) E(\acute{t})},
\end{equation}
where $\tilde{\epsilon}^{(2)}_{BO}$is the second-order permittivity
after the Born-Oppenheimer approximation has been made. The
second-order permittivity $\tilde{\epsilon}^{(2)}$ is given by
$\tilde{\epsilon}^{(2)}= \epsilon_{0} \chi^{(2)}$, where
$\chi^{(2)}$ is the second-order nonlinear susceptibility of the
medium. We want to produce second harmonic in diamond.
 Since, diamond does display inversion
symmetry, $\chi^{(2)}$ vanishes in
 diamond and consequently it can
not produce second harmonic. So, for generation of second harmonic
in diamond, we can use electric field induced second harmonic
generation (EFISHG) to introduce the moving $\chi^{(2)}$ [12].
EFISHG can be described by nonlinear polarization as
\begin{equation}
\mathrm{P^{(2)}=\epsilon_{0}\chi^{(3)} E_{pump} E_{seed}E_{seed}},
\end{equation}
where, $\mathrm{E_{seed}}$ and $\mathrm{E_{pump}}$ are the optical
and applied electric fields, respectively. The moving $\chi^{(2)}$
is given by
\begin{equation}
\mathrm{\chi^{(2)}(z-vt)=\chi^{(3)} E_{pump}},
\end{equation}
where, the third order susceptibility for diamond is
$\mathrm{\chi^{(3)}=0.4 \times 10^{-13}
esu=5.5\times10^{-22}m^2/v^2}$
where,$\mathrm{\chi^{(3)}(SI)=4\pi/(3\times10^{4})^{2}\chi^{(3)}(guassian)}$
[13].
 \\For optimal high harmonic generation, we can
use spatiotemporal quasi-phase matching (QPM) that considers the
case in which momentum is assumed to be conserved at the expense of
an energy mismatch, or where a mismatch in both momentum and energy
is allowed[14].
 The momentum mismatch and the energy mismatch are given by
\begin{equation}
\mathrm{\Delta
k=qk_{0}-k_{q}=q(\frac{n(\omega_{0})\omega_{0}}{c})-\frac{n(\omega_{q})\omega_{q}}{c}},
\end{equation}
\begin{equation}
\mathrm{\Delta\omega=q\omega_{0}-\omega_{q}},
\end{equation}
Where, Eq.(8) leads to
\begin{equation}
\mathrm{\omega_{q}=q\omega_{0}-\Delta\omega}
\end{equation}
The distributed phase mismatch  between the momentum and energy
mismatch can be obtained by using Eqs.(7) and (9),
\begin{eqnarray}
\mathrm{\Delta
k}&=&\mathrm{q(\frac{n(\omega_{0})\omega_{0}}{c})-\frac{n(\omega_{q})(q\omega_{0}-\Delta\omega)}{c}} \nonumber \\
&=&\mathrm{\frac{\Delta\omega
n(\omega_{q})}{c}+\frac{q\omega_{0}[n(\omega_{0})-n(\omega_{q})]}{c}}
\end{eqnarray}
  The QPM condition, describing both quasi-momentum and
quasi-energy conservation, is satisfied if there exists a
$\mathrm{\vec{K}}$ such that $\mathrm{\Delta\vec{K}-\vec{K}=0}$ for
the phase mismatch four-momentum
$\mathrm{\Delta\vec{K}=(\frac{-\Delta\omega}{c}, \Delta k_{x},
\Delta k_{y}, \Delta k_{z})}$. The simple case, only one spatial
coordinate applies:
\begin{equation}
\mathrm{\Delta\vec{K}=(\frac{-\Delta\omega}{c}, \Delta k)},
\end{equation}
\begin{equation}
\mathrm{\vec{K_{1}}=(\frac{-\Omega}{c}, K)},
\end{equation}
Substituting  Eqs. (11) and (12) in
$\mathrm{\Delta\vec{K}-\vec{K}=0}$, leads to
\begin{equation}
\mathrm{\Delta K=K},\ \mathrm{\Delta\omega=\Omega},
\end{equation}
\vskip 1cm {\bf\large 3. Simulations} \vskip 0.5cm In the
simulation, the moving $\chi^{(2)}$ has been defined as
\begin{equation}
\mathrm{\chi^{(2)}=\chi_{eff}^{(2)}g(z-vt)}
\end{equation}
where, $\mathrm{\chi_{eff}^{(2)}=\chi^{(3)}A}$, A is the amplitude
of pump electric field and g(z-vt) is introduced as a normalized
geometrical factor describing the spatiotemporal modulation of the
nonlinear electric susceptibility and given by
\begin{eqnarray}
\mathrm{g(z-vt)}&=&\mathrm{e^{-\frac{(t-\frac{z}{nv}+\frac{n_{center}}{c})^{2}}{n_{sigma}^{2}}}cos(\frac{2\pi
c}{\Lambda}(t-\frac{z}{nv}+\frac{n_{center}}{c}))}\nonumber \\
&=&\mathrm{g(K z-\Omega t)},
\end{eqnarray}
where $n_{center} = \mathrm{60\mu m}$ and $\mathrm{\Omega=2\pi
c/\Lambda}$.  So, the modulation velocity is given by
\begin{equation}
\mathrm{v_{modulation}=v_{phase}=\frac{\Omega}{K}=\frac{\Delta\omega}{\Delta
k}},
\end{equation}
 In order to increase the duration of interaction between seed
electric field and pump electric field, we can choose a large
$n_{sigma}$ or we can use only \textit{cos} term. In the simulation,
it has been seen that there is no q harmonic at
$\mathrm{q\omega_{0}}$ but, there are two harmonic frequencies,
i.e.,
\begin{equation}
\mathrm{\omega_{q}=q\omega_{0}\mp\Delta\omega},
\end{equation}
where, $\Delta\omega=2\pi c/\Lambda$ which is the same as Eq. (13).
Therefore, there are two $\mathrm{\Delta k}$, i.e.,
\begin{equation}
\mathrm{\Delta k=\frac{\pm\Delta\omega
n(\omega_{q})}{c}+\frac{q\omega_{0}[n(\omega_{0})-n(\omega_{q})]}{c}}
\end{equation}
So, the modulation velocities can be written as
\begin{eqnarray}
\mathrm{v_{modulation}}&=&\mathrm{\frac{\Delta\omega}{\mid\Delta k\mid}}\nonumber \\
&=&\mathrm{\frac{\Delta\omega}{\mid\frac{\pm\Delta\omega
n(\omega_{q})}{c}+\frac{q\omega_{0}[n(\omega_{0})-n(\omega_{q})]}{c}\mid}},
\end{eqnarray}
 \vskip 1cm {\bf\large 4. Results and Discussion} \vskip 0.5cm
In this paper, we want to find a $\Lambda$ that is satisfied in
condition of $\mathrm{v_{moving \ \chi^{(2)}}=v_{modulation}}$. For
this purpose, we use the spatiotemporal QPM condition which is
satisfied if there exists a wave vector $\mathrm{\vec{K}}$ such that
$\mathrm{\Delta\vec{K}-\vec{K}=0}$. So, we should find a
$\mathrm{\vec{K}}$ that is equal to $\mathrm{\Delta\vec{K}}$. In one
dimensional (1D) spatial coordinate, we can use Eq.(13). As a
possible experiment, it can be suggested an intense Bessel-Gaussian
beam (which has Bessel spatial profile  and Gaussian temporal
profile) coupled to $\mathrm{\chi^{(3)}}$ and created the moving
$\mathrm{\chi^{(2)}}$, i.e. $\mathrm{\chi^{(2)}=\chi^{(3)}E}$. In
this case, $\Delta\omega=2\pi c/\Lambda$ will just be $\Omega$, (the
Bessel beam frequency, according to eq. (13)). The velocity of
moving $\mathrm{\chi^{(2)}}$ will be
$\mathrm{v_{phase}/cos(\theta)}$,where
$\mathrm{v_{phase}=c/n(\Omega)}$ and $\mathrm{n(\Omega)}$ is
refractive index of diamond. So by changing $\theta$, we can have
any velocity we want to be larger than $\mathrm{v_{phase}}$. The
$\mathrm{\vec{K}}$ related to moving $\mathrm{\chi^{(2)}}$ is given
by $\mathrm{K cos(\theta)}$, where $\mathrm{K=n(\Omega)\Omega/c}$.
At first, we plot variations of
$\mathrm{v_{modulation}=\Delta\omega/\mid\Delta k\mid}$ and velocity
of moving $\mathrm{\chi^{(2)}}$,
$\mathrm{(v=c/n(\Omega)cos(\theta))}$ versus $\mathrm{\Lambda}$ for
fixed $\theta$. By finding intersection for fixed $\theta$, we can
obtain a $\mathrm{\Lambda}$ that is satisfied in condition of
$\mathrm{v_{moving \ \chi^{(2)}}=v_{modulation}}$. For example, in
figures (1) and (2), intersections are obtained for angles 25 and
35, respectively. The numerical results for $0<\theta<90$, are shown
in a graph of $\Lambda$ versus $\theta$ in figure (3). Note that
above results are just for
$\mathrm{\omega_{q}=q\omega_{0}-\Delta\omega}$. One predicte that
 by using pump Bessel-Gaussian  electric field , we can
not obtain any intersection for graphs $\mathrm{v_{moving \
\chi^{(2)}}=v_{modulation}}$ versus $\mathrm{\Lambda}$ for
$\mathrm{\omega_{q}=q\omega_{0}+\Delta\omega}$. We, considered a
seed electric field ($\mathrm{E_{seed}=10^{10} V/m}$) at
$\lambda=1.65 \mu m$ and a pump Bessel-Gaussian electric field
($\mathrm{E_{pump}=5\times10^{8} V/m}$) for angles 25 and 35.
Figures (4) and (5), show the electric field profile and its
spectrum after propagating as far as $\mathrm{1.2 mm}$. Here, the
red profile is pump electric field and the blue one is the seed
electric field. We see the fundamental frequency
($\mathrm{\omega_{0}=1.14\times10^{15} s^{-1}}$) to the left and its
second harmonic (that is seen at
$\mathrm{\omega_{2}=2\omega_{0}\mp\Delta\omega}$ not exactly at
2$\omega_{0}$) to the right. Actually, the spatiotemporal QPM,
shifted the second harmonic to
$\mathrm{\omega_{2}=2\omega_{0}\mp\Delta\omega}$. Since, we have
used spatiotemporal QPM condition at
$\mathrm{\omega_{2}=2\omega_{0}-\Delta\omega}$, the spatiotemporal
QPM  occurs for $\mathrm{\omega_{2}=2\omega_{0}-\Delta\omega}$. We
can clearly see in figures (4) and (5) that, as expected, the second
harmonic components lag behind those of the fundamental, due to
their lower group velocity. In figures (6) and (7), we show the
conversion of energy from the fundamental to the shifted second
harmonic $\mathrm{\omega_{2}}$ for angles 25 and 35, respectively.
The staircase-like features due to quasi-phase matching is just the
fact we expect. For traditional spatial QPM, we have
\begin{equation}
\mathrm{\Delta k_{Q}=q k_{1}-k_{q}-\frac{2\pi}{\Lambda_{spatial
QPM}}=0},
\end{equation}
where, $\mathrm{\Delta k=q k_{1}-k_{q}}$, is momentum phase
mismatch. From Eq. (20), the coherence length is obtained as
\begin{equation}
\mathrm{\Lambda_{spatial QPM}=\frac{2\pi}{q
k_{1}-k_{q}}}\Longrightarrow \mathrm{l_{coh}=\frac{\Lambda_{spatial
QPM}}{2}=\frac{\pi}{q k_{1}-k_{q}}},
\end{equation}
For spatiotemporal QPM, we have
\begin{equation}
\mathrm{\Delta
k-\frac{\Delta\omega}{v_{g}}=\frac{2\pi}{\Lambda_{\xi}
}},
\end{equation}
where, $\mathrm{\Delta k=\Delta\omega n(\omega_{q})/c + q\omega_{0}
[n(\omega_{0})-n(\omega_{q})]/c}$ and $\mathrm{\Delta\omega=q
\omega-\omega_{q}}$. Eq. (22) leads to
\begin{equation}
\mathrm{\Lambda_{\xi}=\frac{2\pi}{\Delta k-\frac{\Delta
\omega}{v_{g}}}}.
\end{equation}

\vskip 1cm {\bf\large 5. Conclusions} \vskip 0.5cm \ In this paper,
we have investigated second harmonic generation (SHG) in the media
such as diamond, where there is no second susceptibility,
$\chi^{(2)}$. So, to obtain second harmonic generation in these
cases, we have used electric field induced second harmonic
generation (EFISHG) to introduce the moving $\chi^{(2)}$. Second
harmonic generation in the moving $\chi^{(2)}$, is verified by
direct PSSD simulations of the nonlinear Maxwell equations.
Spatiotemporal quasi phase matching  have been used for optimal
second harmonic generation. For this purpose, we have looked for a
$\Lambda$ satisfying the condition of $\mathrm{v_{moving \
\chi^{(2)}}=v_{modulation}}$. We have observed the fundamental
frequency ($\mathrm{\omega_{0}=1.14\times10^{15} s^{-1}}$) at the
left and its second harmonic (that is seen at
$\mathrm{\omega_{2}=2\omega_{0}\mp\Delta\omega}$ not exactly at
2$\omega_{0}$) at the right of the spectrum.

 \vskip 1cm {\bf\large
Acknowledgments} \vskip 0.5cm \ The author would like to express her
sincere gratitude to Prof. Daniele Faccio for his expert advice,
constructive discussion and encouragement during preparation of this
manuscript.

\newpage
\begin{figure}
\input{epsf}
\centerline{\epsfysize=7cm \epsffile{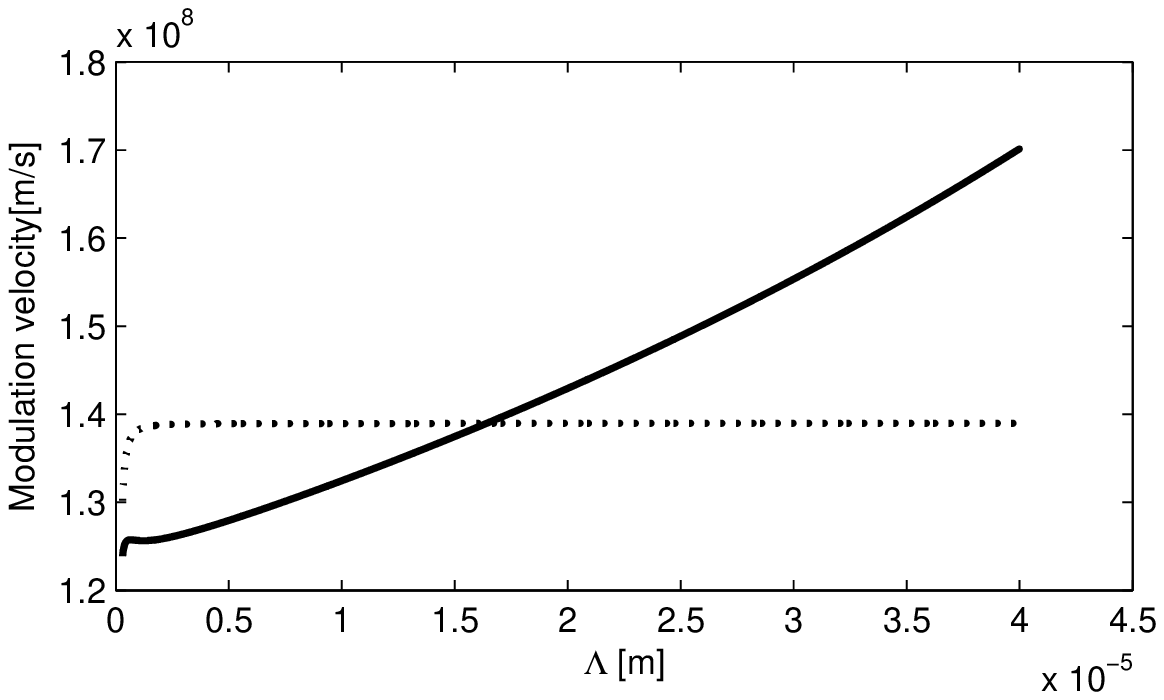}} \caption{
Variations of $\mathrm{v_{modulation}=\Delta\omega/\mid\Delta
k\mid}$ (solid line) and velocity of moving $\mathrm{\chi^{(2)}}$,
$\mathrm{(v=c/n(\Omega)cos(\theta))}$ (dashed line) versus
$\mathrm{\Lambda}$ for $\mathrm{\theta=25}$.}
\end{figure}
\begin{figure}
\input{epsf}
\centerline{\epsfysize=7cm \epsffile{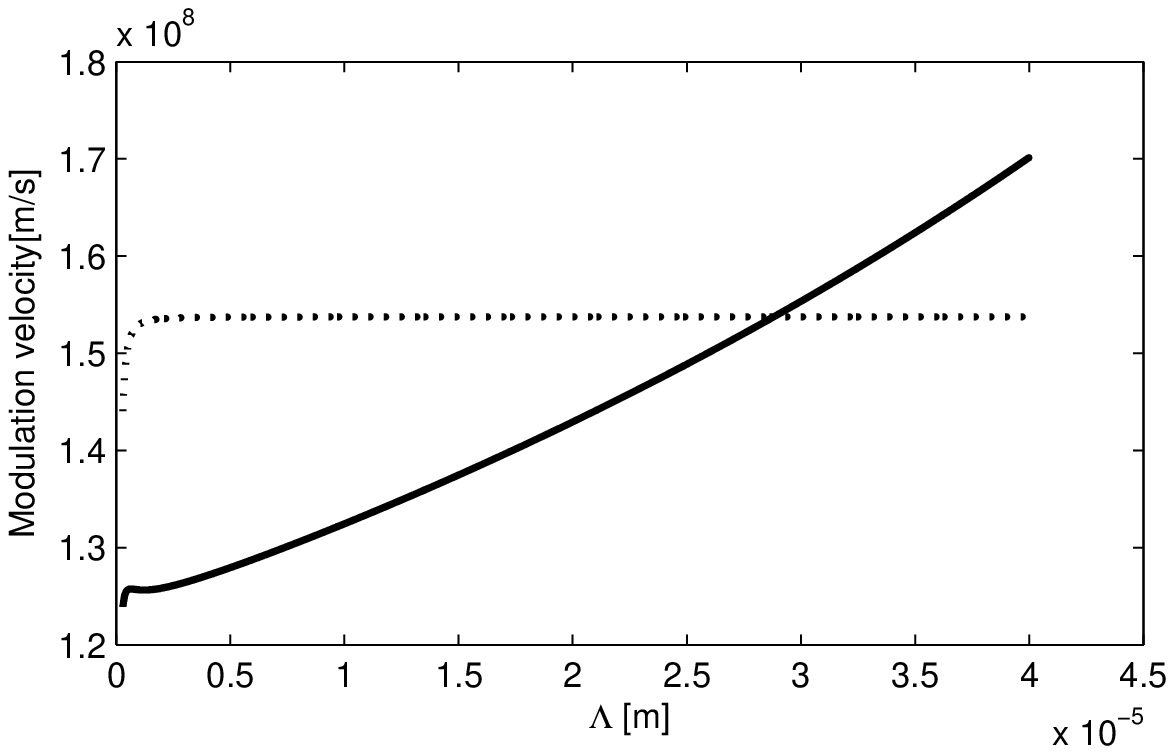}} \caption{Variations
of $\mathrm{v_{modulation}=\Delta\omega/\mid\Delta k\mid}$ (solid
line) and velocity of moving $\mathrm{\chi^{(2)}}$,
$\mathrm{(v=c/n(\Omega)cos(\theta))}$ (dashed line) versus
$\mathrm{\Lambda}$ for $\mathrm{\theta=35}$.}
\end{figure}
\begin{figure}
\input{epsf}
\centerline{\epsfysize=7cm \epsffile{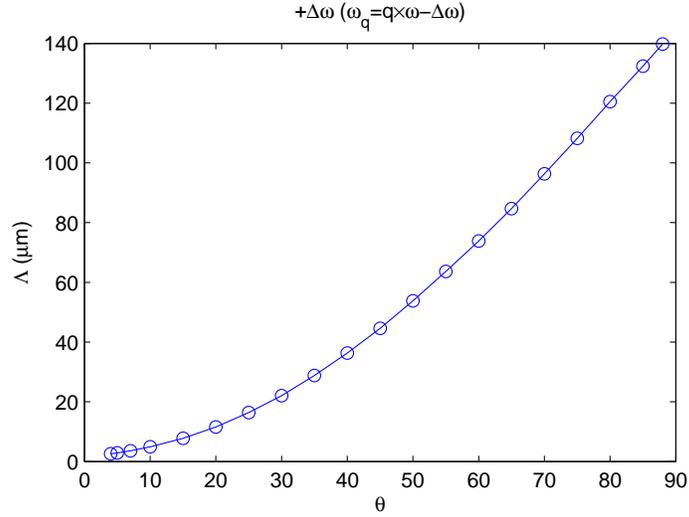}} \caption{Variations
of $\theta$ versus $\Lambda$, which pairs of them are satisfied in
the spatiotemporal QPM condition.}
\end{figure}
\begin{figure}
\input{epsf}
\centerline{\epsfysize=7cm \epsffile{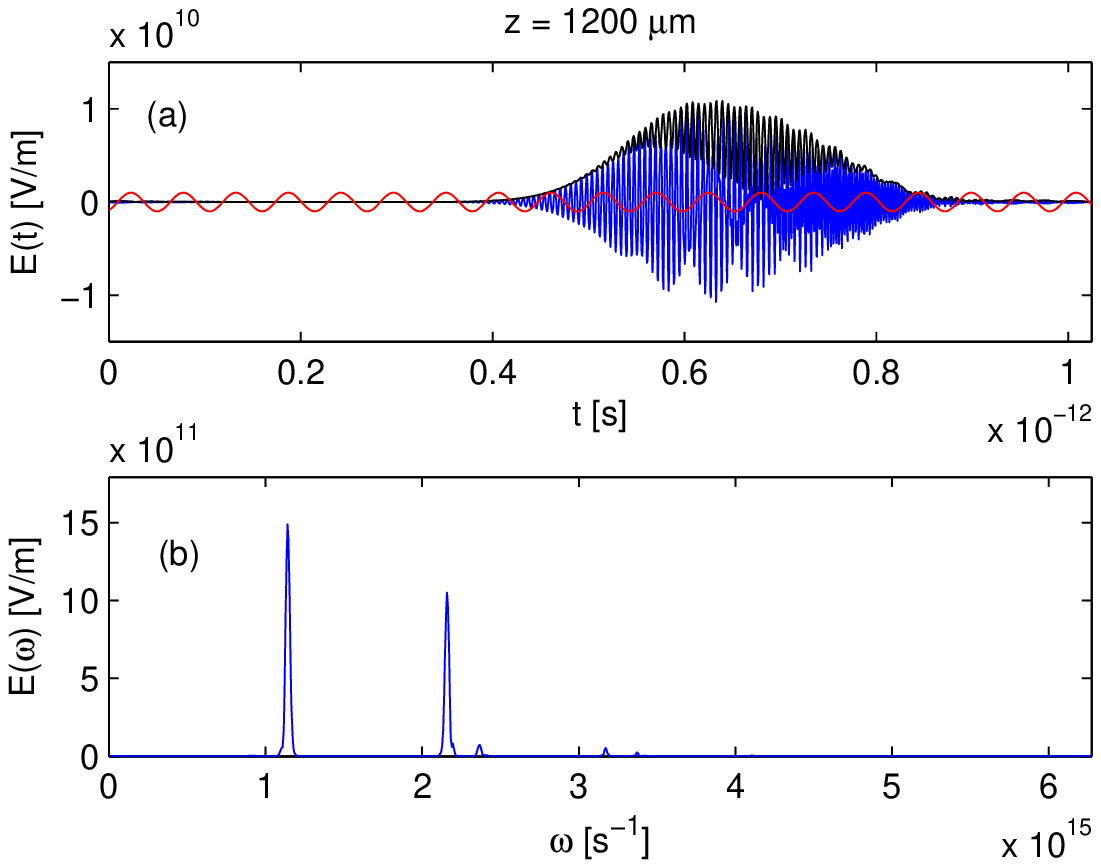}} \caption{(a)The
 electric field profiles of seed (blue)=$10^{10}$ V/m and pump (red)=$5\times10^{8}$ V/m and (b) The seed electric field spectrum  for $\mathrm{\theta=25}$, after
propagating $\mathrm{1.2 mm}$. We clearly see the fundamental
frequency ($\mathrm{\omega_{0}=1.14\times10^{15} s^{-1}}$) to the
left and its second harmonic (that is seen at
$\mathrm{\omega_{2}=2\omega_{0}\mp\Delta\omega}$ not exactly at
2$\omega_{0}$) to the right.}
\end{figure}
\begin{figure}
\input{epsf}
\centerline{\epsfysize=7cm \epsffile{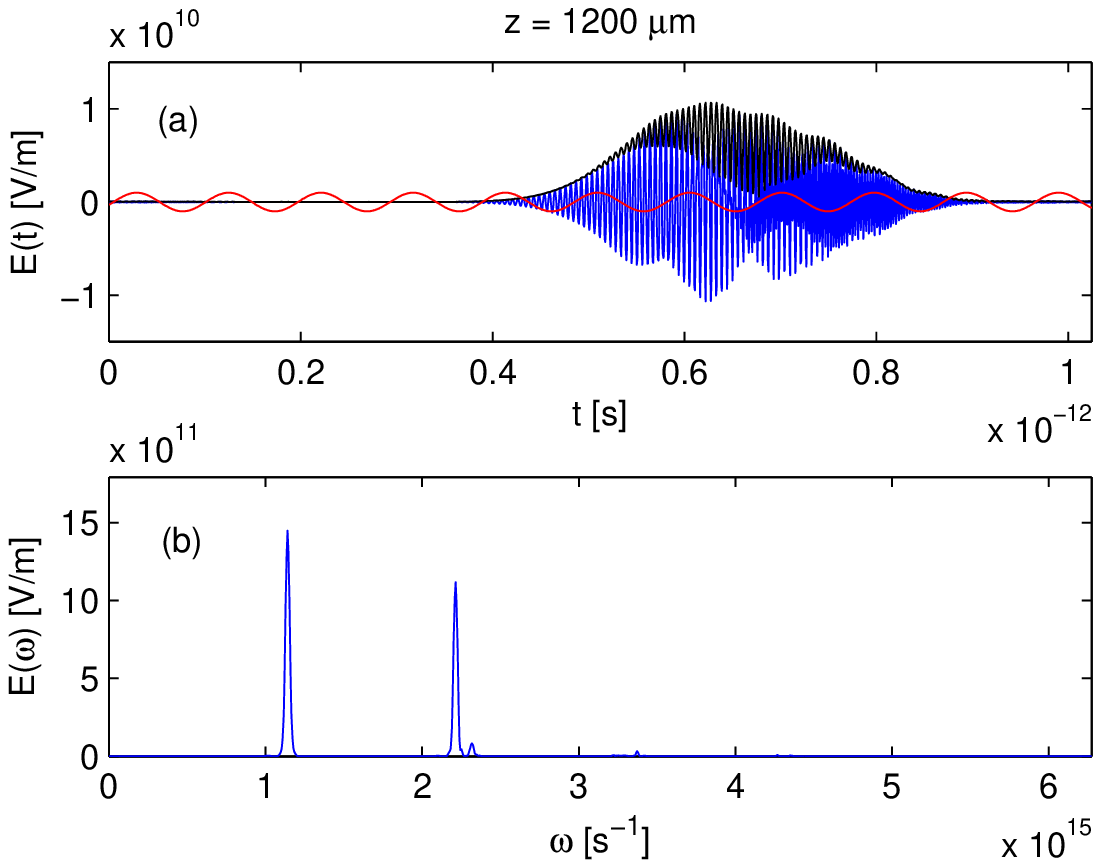}} \caption{(a)The
 electric field profiles of seed $(blue)=10^{10}$ V/m and pump (red)=$5\times10^{8}$ V/m and (b) The seed electric field spectrum  for $\mathrm{\theta=35}$, after
propagating $\mathrm{1.2 mm}$. We clearly see the fundamental
frequency ($\mathrm{\omega_{0}=1.14\times10^{15} s^{-1}}$) to the
left and its second harmonic (that is seen at
$\mathrm{\omega_{2}=2\omega_{0}\mp\Delta\omega}$ not exactly at
2$\omega_{0}$) to the right.}
\end{figure}
\begin{figure}
\input{epsf}
\centerline{\epsfysize=7cm \epsffile{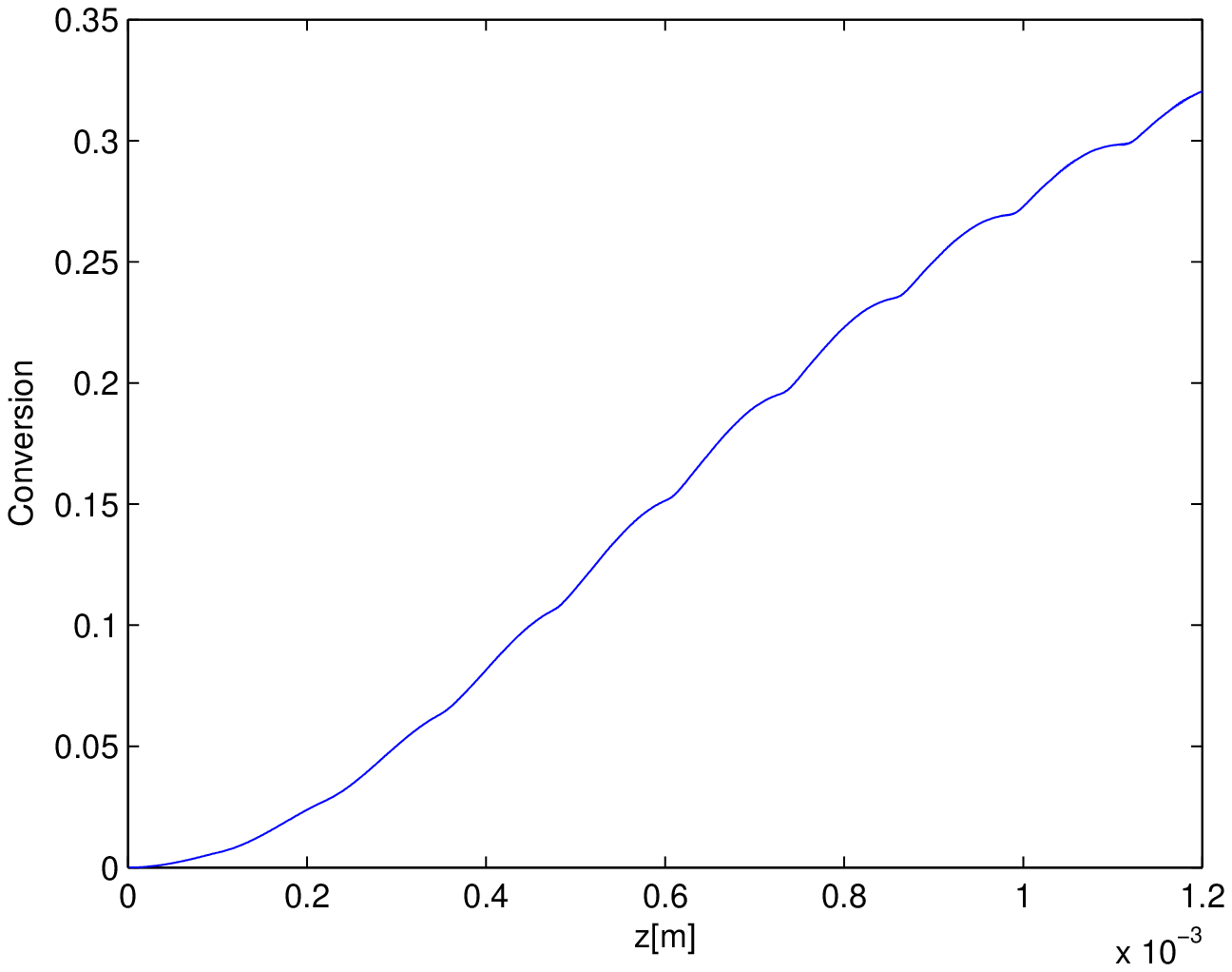}} \caption{conversion
of energy from the fundamental to the shifted second harmonic
$\mathrm{\omega_{2}}$ for $\mathrm{\theta=25}$. The small step-like
features on the plot, are characteristic of QPM.}
\end{figure}
\begin{figure}
\input{epsf}
\centerline{\epsfysize=7cm \epsffile{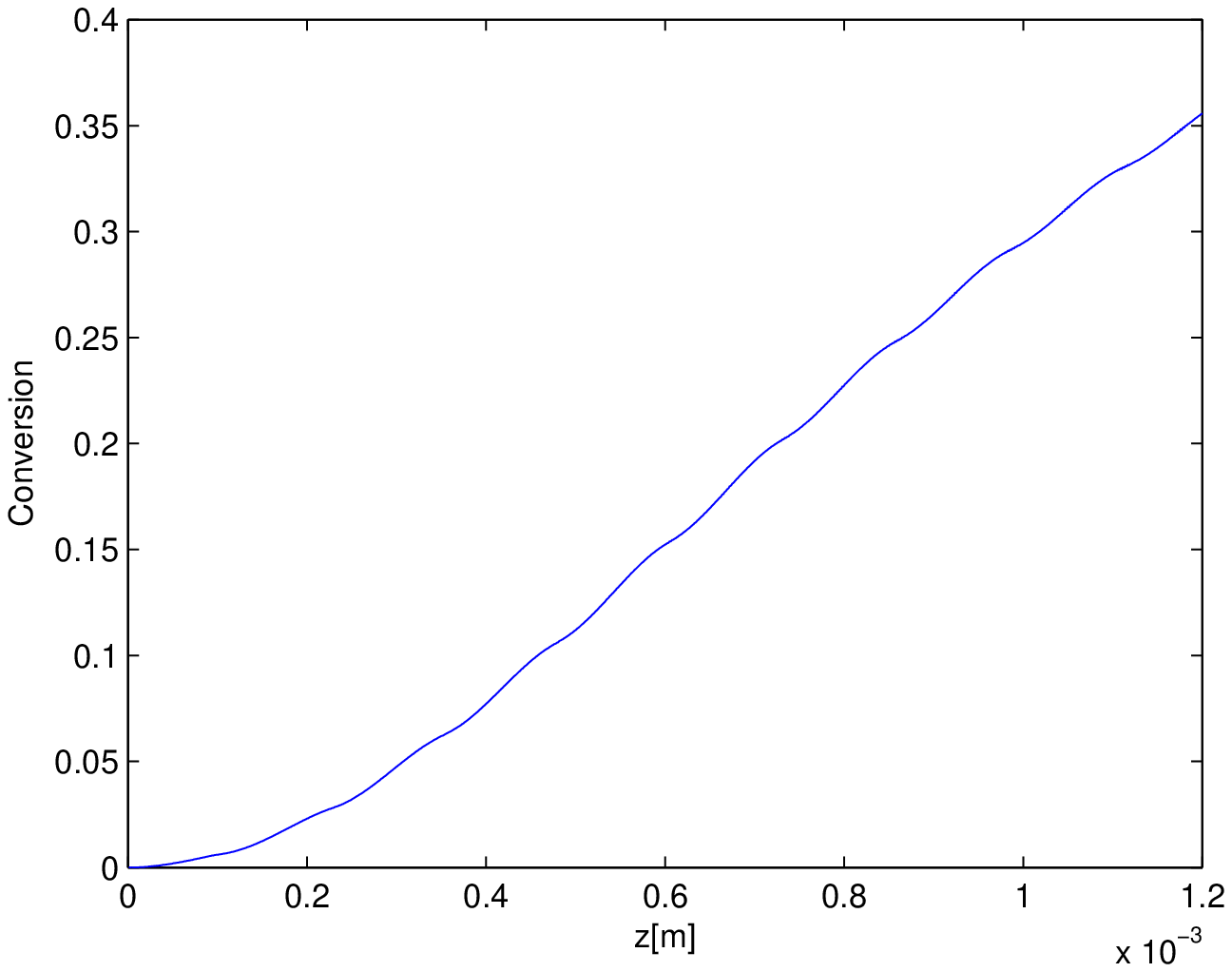}} \caption{conversion
of energy from the fundamental to the shifted second harmonic
$\mathrm{\omega_{2}}$ for $\mathrm{\theta=35}$. The small step-like
features on the plot, are characteristic of QPM.}
\end{figure}

\end{document}